\documentclass[12pt,preprint2]{aastex63}

\newcommand{\kms}{km~s{$^{-1}$}}

\newcommand{\ozone}{He$^{+}$+H$^{+}$}

\newcommand{\nzone}{He$\rm ^{o}$+H$^{+}$}

\newcommand{\tC}{{$ \theta^1$~Ori~C}}

\newcommand{\oiii}{[\ion{O}{3}]}

\newcommand{\Cii}{[\ion{C}{2}]}

\newcommand{\nii}{[\ion{N}{2}]}

\newcommand{\Vscat}{{\bf V$\rm_{scat}$}}
\newcommand{\Vscatnii}{{\bf V$\rm_{scat,[N II]}$}}
\newcommand{\Vscatoiii}{{\bf V$\rm_{scat,[O III]}$}}

\newcommand{\Vcomp}{{\bf V$\rm_{comp}$}}

\newcommand{\Vlow}{{\bf V$\rm _{low}$}}

 \newcommand{\Vpdr}{{\bf V$\rm _{PDR}$}}

\newcommand{\Vmif}{{\bf V$\rm_{mif}$}}
\newcommand{\Vmifnii}{{\bf V$\rm_{mif,[N~II]}$}}

\newcommand{\Vnil}{{\bf V$\rm_{NIL}$}}

\newcommand{\Vlong}{{\bf V$\rm _{long}$}}
\newcommand{\Vlongnii}{{\bf V$\rm _{long,[N~II]}$}}
\newcommand{\Vlongoiii}{{\bf V$\rm _{long,[O~III]}$}}

\newcommand{\Vshort}{{\bf V$\rm _{short}$}}
\newcommand{\Vshortnii}{{\bf V$\rm _{short,[N~II]}$}}
\newcommand{\Vshortoiii}{{\bf V$\rm _{short,[O~III]}$}}

\newcommand{\Slongnii}{{\bf S$\rm_{long,[N~II]}$}}
\newcommand{\Slongoiii}{{\bf S$\rm_{long,[O~III]}$}}

\newcommand{\Snewoiii}{{\bf S$\rm_{new,[O~III]}$}}
\newcommand{\Vnewoiii}{{\bf V$\rm _{new,[O~III]}$}}

\newcommand{\Vomc}{{\bf V$\rm_{OMC}$}}

\newcommand{\Vevap}{{\bf V$\rm_{evap}$}}
\newcommand{\Vevapnii}{{\bf V$\rm_{evap,[N~II]}$}}
\newcommand{\Vevapoiii}{{\bf V$\rm_{evap,[O~III]}$}}

\newcommand{\Sscat}{{\bf S$\rm _{scat}$}}
\newcommand{\Sscatnii}{{\bf S$\rm _{scat,[N~II]}$}}
\newcommand{\Sscatoiii}{{\bf S$\rm _{scat,[O~III]}$}}

\newcommand{\tran}{{\bf SE--NW Transition}}

\newcommand{\inside}{{\bf Inside-Group}}
\newcommand{\outside}{{\bf Outside-Group}}

\newcommand{\lig}{{\bf Low-Ionization-Group}}
\newcommand{\crossing}{{\bf Crossing}}
\newcommand{\cloud}{{\bf Cloud}}

\newcommand{\Scomp}{{\bf S$\rm_{comp}$}}

\newcommand{\Sshortoiii}{{\bf S$\rm_{short,[O~III]}$}}

\begin{document}

\title{Deciphering the 3-D Orion Nebula-II: A low-ionization region of multiple velocity components southwest of \tC\  confounds interpretation of low velocity resolution studies of temperature, density, and abundance}

\author{C. R. O'Dell\affil{1}}
\affil{Department of Physics and Astronomy, Vanderbilt University, Nashville, TN 37235-1807}

\author{N. P. Abel\affil{2}}
\affil{MCGP Department, University of Cincinnati, Clermont College, Batavia, OH, 45103}

\author{G. J. Ferland\affil{3}}
\affil{Department of Physics and Astronomy, University of Kentucky, Lexington, KY 40506}

\begin{abstract}
We establish that there are two velocity systems along lines-of-sight that contribute to the emission-line spectrum of the the brightest parts of the Orion Nebula. These overlie the Orion-S embedded molecular cloud southwest of the dominant ionizing star (\tC). Examination of 10$\times$10\arcsec\ samples of high spectral resolution emission-line spectra of this region reveals it to be of low ionization, with velocities and ionization different from the central part of the Nebula. These properties jeopardize earlier determinations of abundance and physical conditions since they indicate that this region is much more complex than has been assumed in analyzing earlier spectroscopic studies and argue for use of very high spectral resolution or known simple regions in future studies.
\end{abstract}

 \section{Introduction}
 \label{sec:Intro}
 This is the second of a series of papers on the Orion Nebula using high velocity resolution data combined with Hubble Space Telescope imaging. Paper-I dealt with large-scale shells and layers, Paper-III will explore the high spatial resolution spectrum properties of the Orion-S Cloud and the foreground layer of ionized gas, and Paper-IV will be a study of the extended series of shocks forming HH~269. 
The major goal of the present paper is to use groups of spectra in 10$\times$10\arcsec\ samples to understand the large-scale properties of velocity components  as they change across the region southwest of \tC, the dominant ionizing star in the Huygens Region of the Orion Nebula.   We show that this region is fundamentally different from other parts of the Huygens Region, even though its apparent brightness attracts spectroscopic studies to determine characteristics, such as abundances and physical conditions.  The areas we have studied are shown in Figure~\ref{fig:RatioImage}.

\subsection{Background of this study}{\Large }
\label{sec:Background}

In a companion paper \citep{ode20a} (Paper-I) we present an annotated background for the optical studies of the Huygens Region. This is useful to the present study and for brevity is not repeated here. However, the present study focuses on properties to the south-west of the Trapezium stars, including the Orion-S Cloud (henceforth the \cloud ). 

\begin{figure}
 \includegraphics
[width=\columnwidth]
 {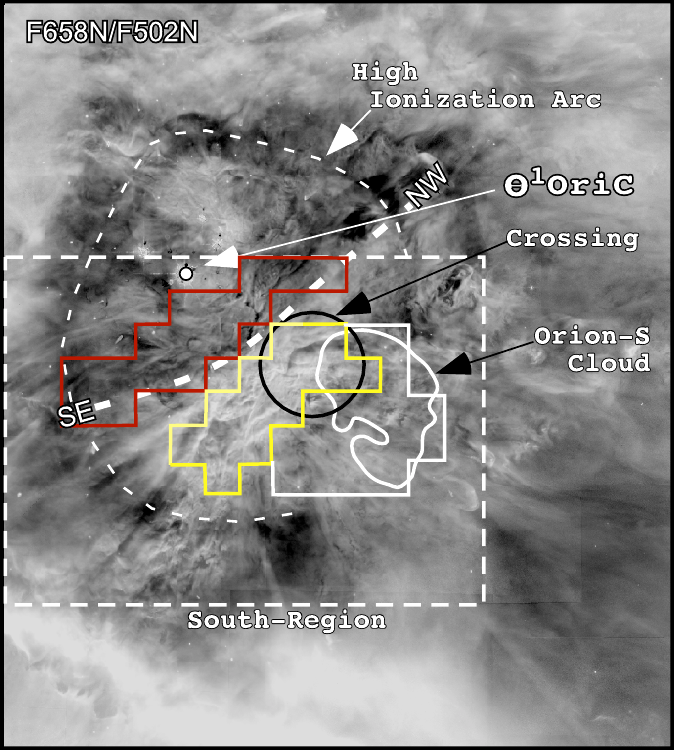}
\caption{This image depicts the signal ratio in the HST F658N and F502N filters (these correspond to the spectrally unresolved lines of \nii\ 658.3 nm and \oiii\ 500.7 nm).
It is a 194\arcsec$\times$216\arcsec\ (0.36$\times$0.40 pc) field of view (FOV) with north up and east to the left and is centered 53\arcsec\ at PA (Position Angle) =237\arcdeg\ from \tC\ and encloses the optically brightest part of the Huygens Region.
The colored lines outline groups repeated and explained in Figure~\ref{fig:SinglePanel}. 
 It also shows the locations of the 21-cm absorption line boundary of the Orion-S Cloud \citep{vdw13}. The dark-line circle indicates the position of the 30\arcsec\ diameter '\crossing ' key region.
The \tran\ is shown with a dashed white line.The dashed white line gives the boundary of the South-Region shown in Figure~\ref{fig:SinglePanel} that is the subject of this study.}
\label{fig:RatioImage}
\end{figure}

We see in Figure~\ref{fig:RatioImage} that the relative signal of the total \nii\ and \oiii\ emission (i.e. line signal without division into velocity components because the narrow-band filters are much wider bandpass than the separation of the components) changes abruptly along a SE to NW line.
 Shown in Figure~\ref{fig:RatioImage}, we designate this feature as the \tran\ where it is depicted as a heavy dashed white line. This feature was first pointed out in \citet{ode09} where it was called the `SE-NW Ionization Boundary'. Given that the NE boundary of the \lig\  lies along the \tran , this indicates that it is a large-scale process that causes the changes, not just what happens in the much smaller \crossing\ region. The \crossing\ Region is of particular interest because multiple stellar outflows originate there and it is studied with higher spatial resolution in Paper-III. 

Several previous studies \citep{ode08,md11,ode18} have established that the region where the \tran\ touches the \crossing\ Region is a nearly edge-on portion of an ionization front in the NE portion of the cloud of material that includes the Orion-S Cloud.  
  
\subsection{Outline of this paper}
\label{sec:outline}
Section~\ref{sec:obs} describes the observational material utilized, the emission-line velocity components, and how they can be interpreted. 
Section~\ref{sec:SouthRegion} presents the results of the deconvolution of the observed velocity components in the targeted region (`the South Region') SW of \tC. 
Section~\ref{sec:conclusions} summarizes the major conclusions of this study.

\subsection{Nomenclature and adopted values}
\label{sec:nomenclature}

The nomenclature and adopted values are listed below.

$\bullet$ {\bf Samples} are areas of 10\arcsec $\times$10\arcsec\ within which spectra from a spatially resolved atlas of spectra of certain emission-lines have been averaged. 

$\bullet$ {\bf Groups} are sets of Samples. 

$\bullet$ \Vshort\ designates a shorter, often weaker,  wavelength velocity component of an emission-line. This was called \Vlow\ in \citet{ode18} and Paper-I.

$\bullet$ \Vlong\ designates the longer, often stronger, wavelength component of an emission-line. This was called \Vmif\ in \citet{ode18} and Paper-I.

$\bullet$ \tran\ is a feature shown in Hubble Space Telescope images that demarques the boundary between the inner and outer portions of the Huygens Region.

$\bullet$ \inside\ is a set of Samples lying on the \tC\ side of the \tran\ feature.

$\bullet$ \lig\ is a set of Samples lying outside of the \tran\ feature.

$\bullet$ \outside\ is a set of Samples lying outside of the \lig\ feature.

$\bullet$ The adopted distance is 388$\pm5$ \citep{mk17}. 

$\bullet$ The adopted velocity for the background PDR is \Vpdr\ = 27.3$\pm$0.3 \kms. 

$\bullet$ All velocities are expressed in \kms\ in the  Heliocentric system (Local Standard of Rest velocities are 18.1 \kms\ less). 

$\bullet$ Directions such as Northeast and Southwest are often expressed in short form as NE and SW.

$\bullet$ In this acronym laden paper acronyms are presented in parentheses following the first use of the word or term being compressed.

\section{Spectroscopic Observations}
 \label{sec:obs}

As in Paper-I we have drawn on the high-spectral-resolution Spectroscopic Atlas of Orion Spectra \citep{gar08} (the Atlas). The Atlas was compiled from a series of north-south spectra at intervals of 2\arcsec\ and have a velocity resolution of 10 \kms.
The resolution along each slit was seeing limited at about 2\arcsec. We have utilized the spectra of \nii\ at 658.3 nm and \oiii\ at 500.7 nm. 
We also employ emission-line images made with the Hubble Space Telescope (the HST) \citep{ode96,ode09} that isolate diagnostically useful emission-lines 
covering the Huygens Region. The regions that we have studied spectroscopically are shown in Figures~\ref{fig:RatioImage} and \ref{fig:SinglePanel}. We have used the previously published results from Paper-I for a North-Region and SE-Region lying on the boundaries of the newly designated South-Region.
    

\subsection{Characteristic Velocity Systems}
\label{sec:VelSys}

In Paper-I (Sections 2.1 and 2.2) we describe how Samples of 10\arcsec $\times$10\arcsec\ were created and de-convolved using 
the IRAF\footnote{IRAF is distributed by the National Optical Astronomy Observatories, which is operated by the Association of Universities for Research in Astronomy, Inc.\ under cooperative agreement with the National Science foundation.} task 'splot'. The emission-line spectra 
show multiple velocity components that were used in \citet{abel19}.  In descending velocity these are \Vscat\ (ascribed to backscattering from dust particles in the background PDR), \Vnewoiii\ (ascribed to material accelerated away from \tC\ by its stellar wind, \Vlong\ (the longer wavelength component, usually ascribed to emission from the ionized layer on the far side of \tC ), and \Vshort\ (a shorter wavelength component usually weak and ascribed to a foreground Nearer Ionized Layer (NIL) lying in the foreground of \tC. When discussed as emission from specific physical layers, the terms \Vmif\ and \Vnil\ are used. These components 
are seen in both the \nii\ and \oiii\ emission-lines. The accuracy of their identification is discussed in Paper-I. The total signal (in instrumental units) are expressed for example as \Slongoiii ). As shown in \citet{ode18} a ratio of S$\rm _{[N~II]}$/S$\rm _{[O~III]}$ = 1.00 corresponds to a calibrated surface brightness ratio (in ergs) of 0.13. 

\subsubsection{Limits on the detection of weak components}
\label{sec:limits}
It is often the case that the spectra are dominated by a single component. We demonstrated in Appendix A of Paper-I that the limits of detection of the weak component  is primarily determined by the Full Width at Half Maximum (FWHM) of the strong component. A rule-of-thumb is that the limit of measurement of the separation is FWHM -0.4 (\kms) for secondary components about 5\%\ the signal of the strong component.  For the groups of data that we use in this study FWHM(\nii) = 17.2$\pm$0.6 \kms\ and 
FWHM(\oiii\ = 15.3$\pm$0.7 \kms , thus setting the limits at slightly less than these numbers. This has led us in most cases to not use the results for \Vshortnii\ and \Vshortoiii , but the \Vscat\ components are strong enough to be retained. When two components are of more similar signal, this limitation does not apply. 

\subsubsection{Expected velocity changes in \Vmif}
\label{sec:Vexpected}

Even if the underlying PDR was a constant velocity (this seems to be true at the level of a few \kms\ according to the
\Cii\ 158 $\mu$m emission mapped by \citet{goi15}) there can be variations in \Vmif\ associated with the tilt of the Main Ionization Front (MIF). If the MIF lies in the plane of the sky, 
one would expect the observed radial velocity to be \Vomc\ - \Vevap , where \Vomc\ is the velocity of the host Orion Molecular Cloud (taken here to be the same as \Vpdr , which is 27.3$\pm$0.3 \kms (\Vomc\ determined from molecules is 25.9$\pm$1.5 \kms\ \citep{ode18}, and \Vevap\ is the rate at which the ionized gas in the MIF evaporates away from the PDR.
In a spectrum of a region that lies exactly perpendicular to the plane of the sky, the \Vmif\ would be the same as \Vomc. The best example of a tilted region of the MIF is the Bright Bar, where spectra show the expected increase in \Vmif\ \citep{ode18}, although even in that case the \Vmif\ values do not reach \Vomc .

\subsubsection{Expected velocity differences of the \Vcomp\ and \Vscat\ components}
\label{sec:VelDif}

In the case of backscattering from a flat-on PDR, which corresponds to the \Vmif\ component being blue-shifted with respect to the PDR, one expects that the difference of the two velocity components 
to be \Vscat -\Vcomp~$\simeq$~2$\times$(\Vpdr -\Vcomp). Since \Vcomp\ = \Vpdr - \Vevap, one expects  \Vscat -\Vcomp~$\simeq$~2$\times$\Vevap. These approximations correspond to the detailed models of extended emission and scattering areas of \citet{hen98}. Later, in \citet{hen05} it was shown that \Vevap\ should be greater for \oiii. Therefore we would expect that 
\Vscat -\Vcomp\ would be greater for \oiii\ than \nii , which is the case (Section~\ref{sec:DeltaVs}). 


Different numbers should apply if one observes a tilted region. As noted in Section~\ref{sec:Vexpected} it is expected that the observed \Vmif\ should increase with increasing tilt as the Line-of-Sight  (LOS) component of \Vevap\ is less, finally reaching \Vpdr\ when the region is seen edge-on. The scattering layer would see the same diminution of the LOS velocity. The expectation would then be that \Vscat -\Vmif\ should go to zero. We cannot test this using the Bright Bar because examination of the slit spectra profiles in \citet{ode18} shows that the \Vscat\ components disappear at the maximum tilt velocities. In any event, we expect that \Vscat -\Vmif\ should decrease with increasing tilt.

\subsubsection{Expected \Sscat /\Scomp\ ratios}
\label{sec:Ratios}

The ratio of signals can be used as a diagnostic. For either backscattering from particles in the PDR lying behind (away from the observer) or from 
a layer of grains in the foreground, \Sscat /\Scomp\ should be much smaller than one. If the \Sscat\ component arises from backscattering, a large value of the ratio would demand that either the albedo is uncharacteristically high or that the scattered light is beamed back towards the source (and the observer), both of which are unlikely, or that the the \Vcomp\ is not producing the light that is scattered.
If the \Sscat\ comes from a foreground layer, a high ratio would indicate a large optical depth in grains, which would in turn mean that the scattering layer is also optically thick to ionizing radiation and the foreground layer would be ionization bounded \citep{ode18}, for which there is no evidence (Paper-I).

 \begin{figure}
\includegraphics
[width=\columnwidth]
 {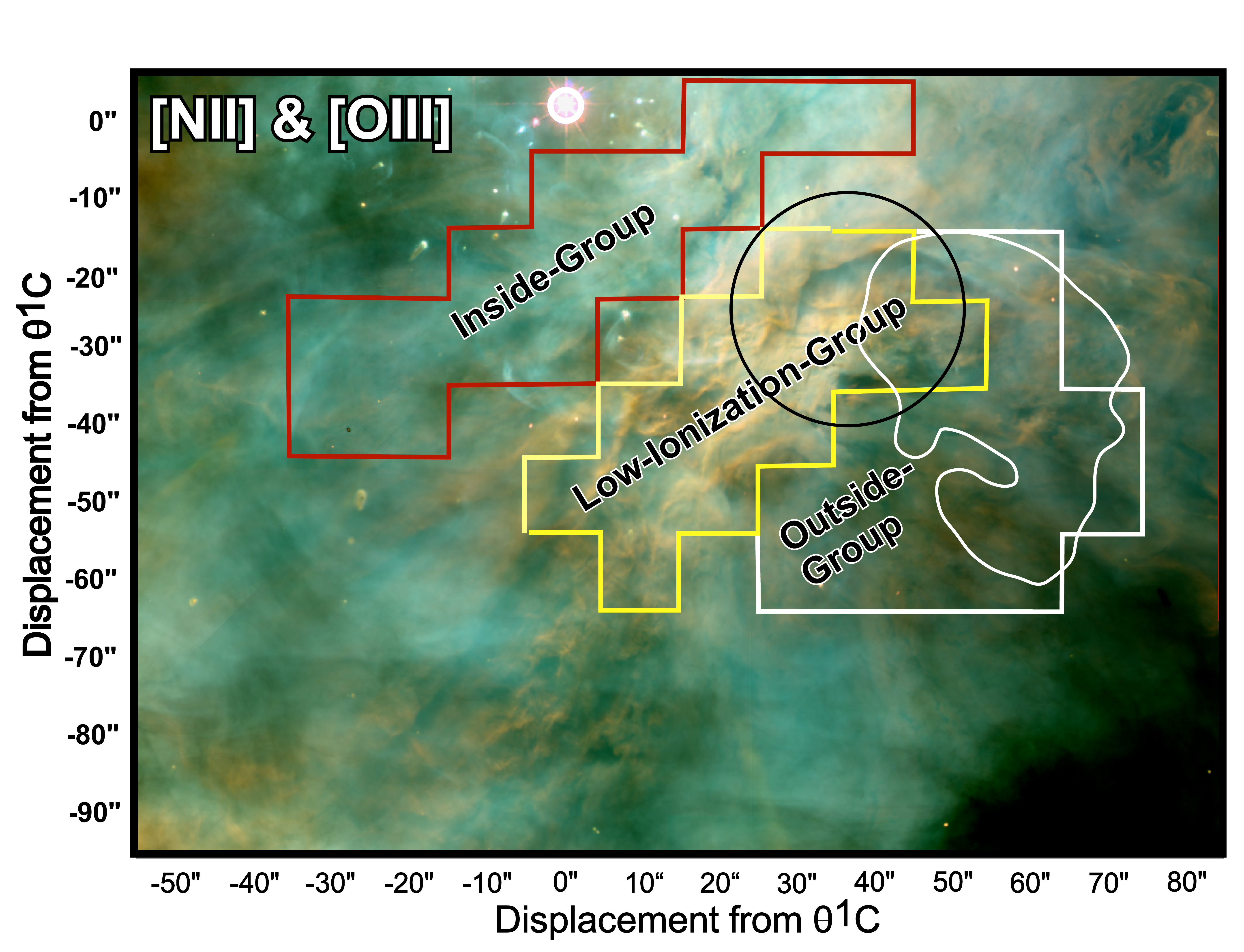}
\caption{This 140$\times$100\arcsec\ image of the South Region is centered 15\arcsec\ west and 45\arcsec\ south of \tC\ and shows some of the same features as in Figure~\ref{fig:RatioImage}.  Displacements south of \tC\ are expressed as negative numbers and displacements west of \tC\ are expressed as positive numbers (in contrast with normal expressions of RA).The Sample spectra were examined in the high S/N ratio 10\arcsec $\times$ 10\arcsec\ samples from spectra in the Atlas. The red border \inside\ samples spectra near \tC , 
the yellow border defines the \lig\ best sampling the raised area on the NE portion of the Orion-S Cloud, and the white border \outside\ samples the region containing the observer's side of the Orion-S Cloud.
}
\label{fig:SinglePanel}
\end{figure}

\section{Properties of the South-Region}
\label{sec:SouthRegion}

In order to examine the properties of the nebula in the region designated in Figures~\ref{fig:RatioImage} and \ref{fig:SinglePanel}, we have identified three  groups of samples within the South-Region. The \inside\ group lies on the \tC\ side of the
\tran\ and samples a region where the MIF is tilted about 15\arcdeg\ \citep{hen05}. The \lig\ is outside the \tran\ and samples a broader region that includes the \crossing . The \outside\ is a further region of overlying material on the observer's side of the Orion-S Cloud. We have determined for these groups the average values of the diagnostically most useful observational parameters and present them in Table~\ref{tab:Groups}. 

\subsection{\nii\ in the South-Region}
\label{sec:propertiesnii}

The \Vlongnii\ values in the several groups all lie within their common uncertainties. This indicates that all three groups represent regions with about  the same tilt. This is not surprising since none of them sample the area along the \tran\ that is known to be highly tilted. 

All of the \Vscatnii -\Vlongnii\  values (16 \kms) for the three groups lie within their uncertainties. 

The low values of \Sscatnii /\Slongnii\ indicate that the \Vscatnii\ components arise from backscattering of the \Vmifnii\ component.

 \begin{deluxetable*}{lccc} 
 \floattable
 \setlength{\tabcolsep}{0.02in}
 \tabletypesize{\scriptsize}
\tablecaption{Properties of Groups*
\label{tab:Groups}}
\tablewidth{0pt}
\tablehead{
\colhead{Discriminator}                     &{\inside}                          & {\lig}              &{\outside}}
\startdata
\Vlongnii                 & 22$\pm$3(15)**                           & 21$\pm$2(12)              &19$\pm$2(15)   \\
\Vshortnii                & 3$\pm$1(10)                                & 6$\pm$3(8)              &3$\pm$3(11) \\
\Vscatnii -\Vlongnii   & 17$\pm$2(14)                             & 16$\pm$1(12)              &16$\pm$1(15) \\
\Sscatnii /\Slongnii & 0.07$\pm$0.05(14)                & 0.08$\pm$0.03(12)     &0.13$\pm$0.04(15) \\
\Vlongoiii                    & 16$\pm$2(15)**                 &13$\pm$2(6)                                            &11$\pm$2(16)\\
\Vshortoiii                   & 3$\pm$1(3)                       &8$\pm$2(8)                                           &3(1)\\
\Vscatoiii -\Vlongoiii   & 18$\pm$4(13)                    &21$\pm$3(6)                                          &24$\pm$2(14)\\
\Vscatoiii -\Vshortoiii &              ---                          & 27$\pm$6(6)                                         & ---               \\
\Sscatoiii /\Slongoiii   & 0.07$\pm$0.02(11)            &0.12$\pm$0.04(7)                               &0.13$\pm$0.07(16)\\
\Sscatoiii /\Sshortoiii &         ---                                &  0.18$\pm$0.09(7)                                     &   ---          \\
\Vnewoiii                 & ---                                               & 22$\pm$2(5)                                                      & 20$\pm$2(2) \\
\Sscatoiii /\Snewoiii  &                 ---                               &0.50$\pm$0.18(4)                                           & 0.39(1)\\
\Slongnii /\Slongoiii    &1.3$\pm$0.1(15)                   &2.9$\pm$0.6(7)                                   & 1.6$\pm$0.2(16) \\
\Slongnii /\Sshortoiii        &              ---                           &2.9$\pm$0.5(6)                                           & --- \\
\enddata

*All velocities are Heliocentric  and in \kms\ (subtract 18.1 for LSR).

**Numbers within parentheses are the number of samples used in the Group.



\end{deluxetable*} 

\subsection{\oiii\ in the South-Region}
\label{sec:propertiesoiii}

\oiii\ emission behaves very differently than \nii\ emission by some samples having \Vshortoiii\ components and no \Vlongoiii\ components, with Table~\ref{tab:Groups} summarizing its properties. We note that \Vlongoiii\ is about 8 \kms\ less than \Vlongnii , which is consistent with the aforementioned expectation that \Vevapoiii\ will be larger than \Vevapnii\ \citep{hen05}. The \Sscatoiii /\Slongoiii\ values indicate that it is the \Vlongoiii\ component that is the source of the backscattered light, but the \Sscatoiii /\Sshortoiii\ ratio in the \lig\ indicates that \Vshortoiii\ contributes in that region.  

\subsection{\Vnewoiii}
\label{sec:Vred}

A weak \oiii\ component is present in a few of the \lig\ and \outside\ samples. Its velocity is similar to \Vlongoiii\ values, but it  should be classified separately because of its very low signal (\Snewoiii ). Its large value of \Sscatoiii /\Snewoiii\ indicates that it is not a contributor to the backscattered light. 


\subsection{Ionization changes across the South-Region}
\label{sec:ionization} 

The ratio of the MIF signals in \nii\ and \oiii\  (\Slongnii /\Slongoiii\ and \Slongnii /\Sshortoiii) is a useful diagnostic of the conditions within the South-Region. 
The values given in Table~\ref{tab:Groups} reflect what is expected as one progresses from the higher ionization \inside\ near \tC\ (ratio 1.3$\pm$0.1). 

Within the \lig\ the high ratios (\Slongnii /\Slongoiii =2.9$\pm$0.6 and  \Slongnii /\Sshortoiii =2.9$\pm$0.5) indicates that this region is of low ionization. It is remarkable that the \outside\ furthest from \tC\ has dropped to an intermediate value. This must provide a guide for the geometry as one progresses across the Orion-S Cloud.

\subsection{Comparison of Predicted and Observed Values of \Vscat -\Vmif}
\label{sec:DeltaVs}

Within the \inside\ the fact that \Vevapnii\ (5$\pm$3 \kms) is smaller than that for \oiii\ (\Vevapoiii\ = 11$\pm$2 \kms) is consistent with theory \citet{hen05}, although the absolute values are probably larger since \citet{hen05} conclude that the region including the \inside\ is about 15\arcdeg\ out of the plane of the sky and thus not all of the evaporation component is seen.

In Table~\ref{tab:DeltaVs} we show a comparison of the predicted and observed velocity differences for both \nii\ and \oiii , using the method described in Section~\ref{sec:Vexpected}.

Because the \nii\ arises from a thin layer immediately on the observer's side of the underlying PDR, we would expect that the predicted and observed separations would be in closest agreement. The essentially constant value of \Vscatnii -\Vmifnii\ (16 \kms ) in each group argues for the three regions having about the same tilt.  The predicted value of \Vscatnii -\Vmifnii\ (average 13$\pm$2 \kms) agrees with this conclusion within their probable errors and the uncertainty due to the crude model that predicts it should be 2$\times$\Vevapnii

Unlike \nii\ the predicted \Vscatoiii -\Vlongoiii\ are consistently larger than the observed values. This is within the range of uncertainty with the crude model that the value should be 2$\times$\Vevapoiii.


\begin{deluxetable*}{lccc} 
 \floattable
 \setlength{\tabcolsep}{0.02in}
 \tabletypesize{\scriptsize}
\tablecaption{Predicted and Observed Values of \Vscat -\Vlong *
\label{tab:DeltaVs}}
\tablewidth{0pt}
\tablehead{
\colhead{Discriminator}                    &{\inside}                     & {\lig}        &{\outside}}
\startdata
Observed \Vlongnii                    & 22$\pm$3          &21$\pm$2                    &19$\pm$2\\
Observed  \Vscatnii -\Vlongnii   & 17$\pm$2        &16$\pm$1                       & 16$\pm$1\\
Predicted \Vscatnii -\Vlongnii   & 10$\pm$6           &12$\pm$4                       &16$\pm$4\\
Observed \Vlongoiii                    & 16$\pm$2           &13$\pm$2                        &11$\pm$2\\
Observed \Vscatoiii -\Vlongoiii   & 18$\pm$4             &21$\pm$3                & 24$\pm$2\\
Predicted \Vscatoiii -\Vlongoiii   & 22$\pm$4            &27$\pm$3                   &32$\pm$8\\
Observed \Vshortoiii                   &      ---                 & 8$\pm$2                         & --- \\
Observed \Vscatoiii -\Vshortoiii  &  ---                       &27$\pm$6                      & --- \\
Predicted \Vscatoiii -\Vshortoiii   & ---                        &36$\pm$4                     &    --- \\
\enddata

*All velocities are Heliocentric  and in \kms\ (subtract 18.1 for LSR).



\end{deluxetable*} 

\subsection{Comparison with an earlier study}
\label{sec:comparison} 

The South-Region overlaps noticeably with portions of the nebula studied in \citet{ode18}. 
 In the present study we have used the large scale changes in ionization within the South-Region to identify our data samples, whereas
 in the earlier study the sample selection was driven primarily by proximity to the \crossing. 
 
 The most similar samples with the present study are the \lig\ and the SW region of \citet{ode18}, whose results are given in his Table~4.
 A comparison with Table~\ref{tab:Groups} shows significant differences only in the \Vshort\ components, which can be attributed to the higher signal-to-noise ratio data used in the current paper. 
 
The improved method of selection of the samples and their better signal-to-noise ratio means that the current results are to be preferred. Other
regions studied in \citet{ode18} are not addressed in either Paper-I or in this study, therefore, \citet{ode18} remains the best source of information on them. 

\section{Conclusions}
\label{sec:conclusions}

$\bullet$ {\bf Two Velocity systems are recognized.} \Vshort\ is usually associated with foreground layer of blue-shifted ionized gas lying between
\tC\ and the outer, predominantly neutral Veil. \Vlong\ is associated with ionized gas lying on the observer's side of the nebula's Main Ionization Front or 
the observer's side of the \cloud. 

$\bullet$ {\bf The \Vshortoiii\ system }is usually weak as compared with the \Vlongoiii\ system near the \tC , but becomes the dominant 
velocity component as the line-of-sight crosses the \cloud .

$\bullet$ {\bf The \Vlongnii\ values indicate that this emission comes from nearly flat-on regions.}  They are nearly constant, which would indicate that one is viewing samples of the same orientation as one crosses from the sub-\tC\ direction, across the \cloud, and over the main body of the \cloud.

$\bullet$ The region called here the \lig\ differs significantly from other regions in the inner Orion Nebula. It is of lower ionization when comparing \nii\ and \oiii\ surface brightness and its strongest \oiii\ components fall into two distinct velocity groups. It contains the peculiar region called the \crossing . The unusual features of that region are shared by many, but not all the Samples within the \lig . The fact that most of this group lie beyond the NE edge of the Orion-S Cloud indicates that the conditions on the observer's side of the Cloud determine the conditions of this region.

$\bullet$ {\bf A caution on Spectroscopy in the Brightest Parts of the Huygens Region.}
In order to optimize the signal from spectra in the Huygens Region, most studies have been of the brightest parts of this region. Unfortunately, these brightest regions occur in the complex structure where the transition from \oiii\ dominant to \nii\ dominant occurs. Near \tC\ the structure of the nebula is simple, with the \Vlong\ components being dominant. This means that analysis according to single but related \nzone\ and \ozone\ layers should yield valid values of the physical conditions and abundances. However, as one moves into the brightest parts (this occurs as the line-of-sight crosses the \cloud), the dominance of the \Vlongnii\ component remains the same while the \Vshortoiii\ component becomes dominant. This indicates that we are looking through two widely separated regions with an ill-defined link. One should do separate analyses of the \Vshort\ and \Vlong\ components, even if one of them is not the strongest component.

Studies have been made at the necessary high velocity resolution. For example \citet{gar07} used sufficient velocity resolution, but worked with the total line signals and \citet{md09}) used 10 \kms\ resolution over a wide range of wavelengths, but used their data only for exploring the conditions in HH~202.

In addition, because of scattered light from 
the Trapezium stars, ground-based telescope spectroscopic studies have avoided their immediate vicinity. These selection effects mean that it would be wise to examine the results in most of the major studies \citep{bal91,jack00,bla07,md08,md11,rubin03} in the light of the 3-D model that now applies and to make new studies at high velocity resolution of carefully chosen regions within the South-Region, in particular in our \inside .

$\bullet$ The questions raised about the \lig\ in this study call for study at higher spatial resolution, which is the subject of Paper-III of this series.

 \section*{acknowledgements}
 
 We are grateful to Cornelia Pabst of the Leiden Observatory and Pedro Salas of the National Radio Astronomy Observatory for discussions of the distribution of emission from inside the PDR. 
 
The observational data were obtained from observations with the NASA/ESA Hubble Space Telescope,
obtained at the Space Telescope Science Institute (GO 12543), which is operated by
the Association of Universities for Research in Astronomy, Inc., under
NASA Contract No. NAS 5-26555; the Kitt Peak National Observatory and the Cerro Tololo Interamerican Observatory operated by the Association of Universities for Research in Astronomy, Inc., under cooperative agreement with the National Science Foundation; and the San Pedro M\'artir Observatory operated by the Universidad Nacional Aut\'onoma de M\'exico. 
We have made extensive use of the SIMBAD data base, operated at CDS, Strasbourg, France and its mirror site at Harvard University, and NASA's Astrophysics Data System Bibliographic Services. 

GJF acknowledges support by NSF (1816537, 1910687), NASA (ATP 17-ATP17-0141), and STScI (HST-AR- 15018).


\end{document}